\documentstyle[11pt,newpasp,twoside,epsf]{article}
\def\edcomment#1{\iffalse\marginpar{\raggedright\sl#1\/}\else\relax\fi}
\reversemarginpar

\begin{document}
\title{A sub-mm imaging survey of ultracompact HII regions}
 \author{Mark Thompson}
\affil{Centre for Astrophysics \& Planetary Science, School of Physical
Sciences, University of Kent, Canterbury,  UK}
\author{Jenny Hatchell}
\affil{Max Planck Institut f\"ur Radioastronomie, Bonn, Germany}
\author{Geoff Macdonald}
\affil{Centre for Astrophysics \& Planetary Science, Department of Electronics, 
University of Kent, Canterbury, Kent, UK}
\author{Tom Millar}
\affil{Astrophysics Group, Department of Physics, UMIST, Manchester, UK}
\begin{abstract}
We present the preliminary results of a sub-mm imaging survey of ultracompact
HII regions, conducted with the SCUBA bolometer array on the JCMT.
\end{abstract}

\section{Introduction}
Ultracompact (UC) HII regions are currently the best known 
tracer of massive YSOs and represent the earliest confirmed stage of massive
star formation. In excess of 150 UC HII regions have been
detected, mainly by radio surveys. 
Whilst the environments of UC HII regions are known very well on the small scale
(a few arcseconds) they are not well known on scales over 40\arcsec. This is
because most UC HIIs have, to date, been observed using either interferometers
(to gain information on small scales at the expense of large scales) or by
single-position large-beam (typically 40\arcsec~or worse) spectroscopy. To
redress this issue we recently undertook an imaging survey of over 100 UC HII regions
using SCUBA on the JCMT, which enables us to rapidly map (with high resolution) the dust emission from
the clumps in which the UC HIIs are embedded.

\section{The survey}

SCUBA is mainly comprised of two bolometer arrays which (almost) instantaneously
sample a 2\arcmin~field simultaneously at 450 and 850 $\mu$m. The instrument is
very sensitive, with the results that we were able to image each UC HII down to
a 1$\sigma$ noise level of typically 50 mJy at 850 $\mu$m in only 3 minutes. Our
source sample was drawn from the UC HII region catalogues of Wood \& Churchwell
(1989) and Kurtz, Churchwell \& Wood (1994), comprising some 140 UC HII regions
in all. Our motivations for the survey were to \emph{i)} investigate the
large-scale structure of the dust clumps embedding the UC HIIs; \emph{ii)}
fill the existing gap in the spectral energy distribution at sub-mm wavelengths;
\emph{iii)} search for other unknown dust clumps in the field of view, possibly
harbouring massive YSOs or protostars in different evolutionary states and 
\emph{iv)} identify hot molecular cores via their strongly peaked sub-mm
emission (Hatchell et al.~2000).

\section{Preliminary results}

In total we observed 106 out of the 140 UC HII regions in the catalogues. Sub-mm
emission was detected in 80$\%$ of the sample. The morphology of the dust clumps
is surprisingly linear: over half of the singly-peaked clumps are elongated
along one axis (e.g.~G10.84 in Fig.~1) and over half of multiply-peaked clumps
take the form of cores string along a ridge (e.g.~G23.96 in Fig.~1). We also
identified a large number of previously unknown dust clumps in the field of
view, many of which are not associated with radio continuum or embedded IR
sources and may contain
massive protostars in an earlier stage to that of the UC HII (see Gibb et al.,
this proceedings for details of BIMA follow-up observations). In addition we
identified 15 UC HIIs associated with strongly peaked sub-mm continuum  which may
contain hot molecular cores.

%We are following up our SCUBA survey with a programme of molecular line
%observations to confirm the newly detected radio-quiet clumps as star formation
%regions and the strongly-peaked objects as hot cores.
\begin{figure}
%\plottwo{g1907_850.ps}{g2396_850.ps}
\plotfiddle{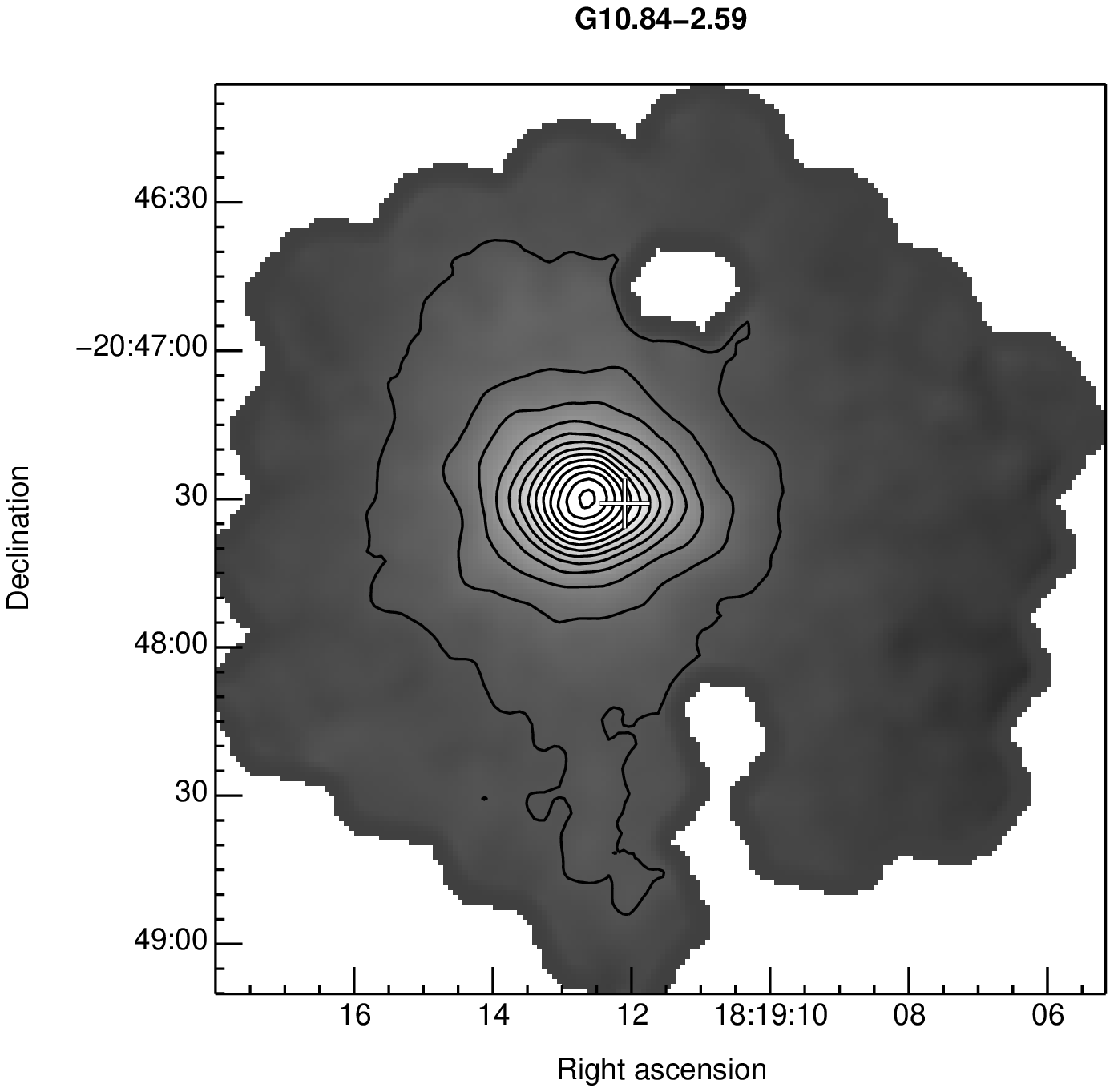}{6cm}{0}{40}{40}{-240}{-70}
\plotfiddle{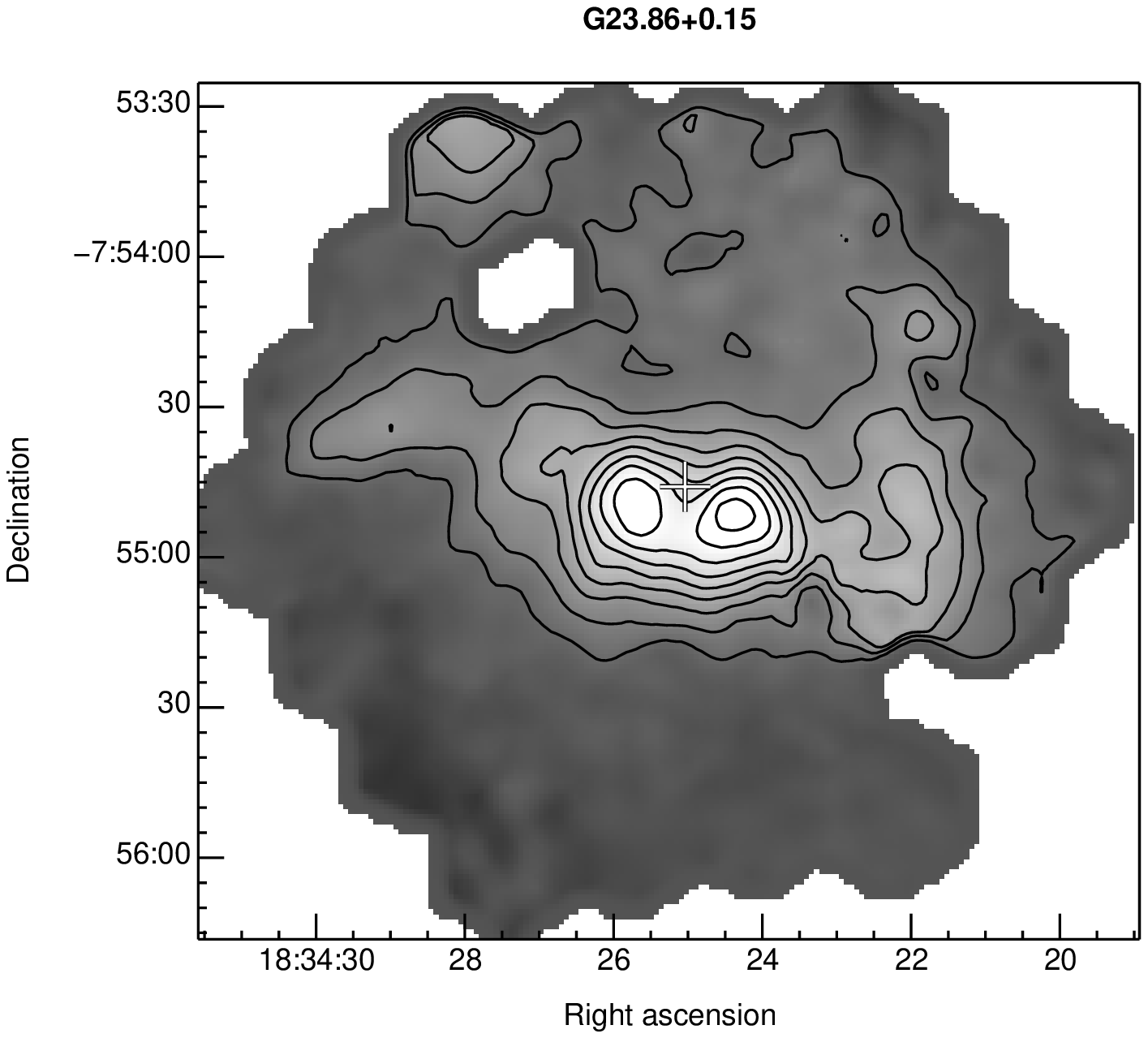}{0cm}{0}{40}{40}{-10}{-50}
%{g2396_850_bw.eps}
\vspace*{-1cm}
\caption{850 $\mu$m images from the survey. Crosses mark the locations of the UC
HII regions. Holes in the image are removed noisy pixels.}
\end{figure}

\end{document}